\documentclass[submission]{eptcs}
\providecommand{\event}{FESCA 2013} 
\usepackage{breakurl}             
\usepackage{latexsym} 
\usepackage{ae}
\usepackage{amsmath}
\usepackage{amssymb}
\usepackage{graphicx} 
\usepackage{booktabs} 

\title{Towards a Formal Framework for \\ Mobile, Service-Oriented Sensor-Actuator Networks}
\author{Helena Gruhn
\institute{Technische Universit\"{a}t Berlin\\ Berlin, Germany}
\email{helena.gruhn@tu-berlin.de}
\and
Sabine Glesner
\institute{Technische Universit\"{a}t Berlin\\ Berlin, Germany}
\email{\quad sabine.glesner@tu-berlin.de}
}
\def\titlerunning{Design Framework for SOSANETs}
\def\authorrunning{Helena Gruhn \& Sabine Glesner}
\begin{document}
\maketitle

\begin{abstract}
Service-oriented sensor-actuator networks (SOSANETs) are deployed in health-critical applications like patient monitoring and have to fulfill strong safety requirements. However, a framework for the rigorous formal modeling and analysis of SOSANETs does not exist. In particular, there is currently no support for the verification of correct network behavior after node failure or loss/addition of communication links. To overcome this problem, we propose a formal framework for SOSANETs. The main idea is to base our framework on the $\pi$-calculus, a formally defined, compositional and well-established formalism. We choose KLAIM, an existing formal language based on the $\pi$-calculus as the foundation for our framework. With that, we are able to formally model SOSANETs with possible topology changes and network failures. This provides the basis for our future work on prediction, analysis and verification of the network behavior of these systems. Furthermore, we illustrate the real-life applicability of this approach by modeling and extending a use case scenario from the medical domain.
\end{abstract}

\section{Introduction}
Mobility and independence are some of the major factors defining the quality of life. With rising age people often develop physical and mental diseases of different degree influencing their options for an independent and mobile life style. Service-oriented sensor-actuator networks (SOSANETs) \cite{Rezgui2007SOSANET} employed as assistant systems at home and en route can support the user by securing their everyday life. These networks are highly dynamic. New communication channels are built ad hoc, connecting the system with other previously unknown SOSANETs. These systems are health-critical and have to fulfill strong safety requirements. Up to now there exists no formal design framework for the verification and analysis of SOSANETs. The common approach is to evaluate network behavior by running test scenarios on prototypes or simulations. These tests do not cover the whole state space of the network and cannot ensure correct system behavior.\\
In this paper, we address the problem of modeling dynamic SOSANETs. We require a formal modeling framework which offers techniques to model topology changes like the establishing of new connections or the sudden disappearance of a component. Especially the interaction with components, which were unknown at design level, has to be representable. Furthermore, we require the framework to offer features for modeling basic functionalities of service-oriented architectures like service publishing. Additionally, the framework needs to be platform-independent to allow the representation of all kinds of nodes.  \\
Our idea is to base our framework on the $\pi$-calculus \cite{pi}. The $\pi$-calculus is a well-established formalism for the modeling of compositional and concurrent systems. It provides primitives for the description and analysis of distributed infrastructures. Furthermore, it offers the modeling foundation for the verification of correct network behavior after topology changes.\\
We choose KLAIM \cite{klaim2003}, a formal language based on the $\pi$-calculus, as a foundation for our approach. It enriches the formalism by providing concrete actions for the manipulation of data within the network. In this paper we evaluate the applicability of KLAIM, which was developed for Wide Area Networks (WAN), for the modeling of SOSANETs. We consider the usability of the privileged node coordinator processes provided within KLAIM for the modeling of anticipated and unexpected topology changes in service-oriented sensor-actuator networks. Additionally, we assess the language to judge its ability to represent basic functions of a service-oriented network. These are the service publishing,  the service discover and the service request. Furthermore, we validate our observations by modeling an use case from the medical domain.\\
The mentioned case study is a driver assistant system for the support of elderly driver. We extend the existing closed system to an open network to match our requirements. The open net is capable of interacting dynamically with expected new network components, e.g. portable medical devices like a pulse meter, and completely unknown networks, e.g. driver assistant systems installed in cars of different brands.\\

The remainer of this paper is structured as follows. In Section \ref{relatedWork} we discuss related work in the area of formal modeling and verification of sensor-actuator networks and other distributed systems. This is followed by an introduction to SOSANETs, the $\pi$-calculus and KLAIM, the basis of our design framework, in Section \ref{Background}. Section \ref{model} illustrates our use case scenario and its proposed extension to an open network. We present a model of the driver assistant system and show the possibilities and limitations of KLAIM as a modeling language for SOSANETs in its current version. Afterwards, Section \ref{conclusion} concludes this paper and gives an outlook on future work.

\section{Related work}\label{relatedWork}

In this section we provide a discussion on literature related to our work. There is no formal modeling framework for SOSANETs to the best of our knowledge. Still, work on formal modeling and verification on  Wireless Sensor Networks (WSN), Sensor-Actuator Networks (SANET), Wide Area Networks and  distributed systems exist, which is related to our application field.\\
\\
Today, it is common practice to check the behavior of WSNs and SANETs by running tests on a simulation or prototype of the network (e.g. \cite{greatduckisland},\cite{Rezgui2007SOSANET}). However, the verification of these networks is an emerging field of research. Some papers were published in recent years presenting approaches for the formal modeling and verification of WSNs, SANETs and other distributed systems.\\
Martinez and al. \cite{martinez2011} present an approach which uses Colored Petri Nets (CPN) for the modeling of SANETs. Their design method allows the integration of different node architecture levels (CPU, operating system, middleware...). They make use of well-known Petri nets properties to verify the network behavior. The biggest drawback of their approach for our task is that they do not consider topology changes at all. Even the break-down of a wireless link, a common event in wireless SANETs, is not representable.\\ 
UM-RTCOM \cite{Diaz2006} is another modeling framework for SANETs. It is written in CORBA and is real-time component based. The communication via a channel is modeled as a tuple. The model can be used to analyze e.g. deadlock freedom or verify liveness properties. UM-RTCOM uses components to capture the system behavior. The model declares for each component interfaces, the services which are offered by each component, and the connectivity of each component. The components can invoke each other based on event triggers, time triggers, or service requests \cite{Jacoub2012}. UM-RTCOM biggest shortcoming for our goal is the absence of tools for handling topology changes.\\
Dearle et al. present the modeling language Insense\cite{dearle2008},\cite{balasubramaniam2008}. It uses a component-based model for WSNs. Components are concurrent and use synchronous communication. They can represent software or hardware entities of a WSN. They capsulate a specific behavior and have interfaces to interact with other components of the system. There are no dependencies between components. Insense supports topology changes. Channels can be connected or disconnected. New component instances may be dynamically created and executing instances may be stopped. This allows the rewiring and replacing of arbitrary components at runtime. In a first approach to show the correctness of Insense, Sharma et al. used Promela constructs for modeling and Spin for verification \cite{sharma2009}. A big misfit with our requirements is that Insense is built on the Contiki operating system \cite{dunkels2004contiki} making the whole modeling procedure platform dependent.\\   
The Promela Model, described by Oleshchuk \cite{oleshchuk2003}, was created to verify the following correctness properties in Spin: sensors are always connected and at least one communication way for each sensor exist. It is assumed that nodes are in a dynamic movement state resulting in unreliable communication channels. The routing protocol ensures that each node is aware of neighboring nodes. The physical location of a node is regularly updated to a central unit, called Location Manager. There is no central unit in our approach due to strong resource restrictions on mobile devices which are the core units of our mobile patient monitoring system.\\
Francalanza and Hennessy present in \cite{francalanza2008} an approach to model node and link failure in distributed programs. They use the distributed $\pi$-calculus framework and extend it by a ping construct for detecting and reacting on those failures. Their approach does not cover the dynamical addition of nodes during run-time and is only applicable to SOSANETs via abstract constructs describing detailed process behavior.
KLAIM, a formal language first published by de Nicola et al. \cite{deNicola1998_klaim}, was developed for modeling Wide Area Networks. A detailed description of KLAIM can be found in Section \ref{Background}. Although it is possible to model topology changes with this language, the addition of unknown (at deployment time) components can not be represented due to the login/accept constructs used for adding new network nodes.\\
None of the described approaches fulfills all our requirements for a formal design framework for mobile service-oriented sensor-actuator networks.

\section{Background}\label{Background}

Our approach is based on work introduced in this section. First, we present SOSANETs explaining the differences to and advantages over general wireless sensor-actuator networks. Afterwards we give an introduction to KLAIM. 


\subsection{Service-oriented sensor-actuator networks} \label{SOSANET}

Sensor-actuator networks (SANETs) consist of nodes with sensing and/or actuation capabilities. Sensors observe their surroundings and collect data. Actuators make decisions and affect the environment by performing various actions \cite{Akyildiz2004}. The communication between nodes can be physical or wireless. Nodes may be stationary or mobile.\\ 
There are two widely-used SANET architectures. Application-specific SANETs are often developed to suit a highly-specialized utilization. These are greatly optimized networks with restricted reusability value due to limited resources. A more open approach is used for generic SANETs, which are not intended for a specific employment but require a generic code installation on every node. This results in unusable code on some nodes, e.g. actuator coordination on a node without an actuator, in spite of limited storage capabilities. These design limitations are reduced by using a service-oriented approach for SANET development. \\
Originally the term \textit{service-oriented architecture} (SOA) refers to a logical set that consists of several large software component that together perform a task or a \textit{service}. The paradigm is popular e.g. in the web software developers world (Web Services) \cite{Papazoglou2003}. However, it can be easily adapted for SANETs by including not only large complex services, but also simple ones. These could be data storage, routing or sensor reading.\\
Service-oriented sensor-actuator networks (SOSANETs) \cite{Rezgui2007SOSANET} realize this concept: every node publishes its own capabilities as a service. A service is a computational component with a unique network-wide identifier. It can be invoked asynchronously and more than one instance can be found within the whole net. Clients can invoke a service through queries which are either directed to a base station or to a specific node. There are two query types: task queries (e.g. reading of a sensor and returning the results) or event queries (e.g. request of a notifier message if an event occured).  \\
The main benefit of this architecture is that the services can be loosely coupled. Only the interface has to be known to provoke a service which allows interaction of devices of diverse producers without revealing construction details. Complex services can be composed using independent SOSANETs through dynamic search and connectivity. This leads to high reusability rates of code and hardware reducing the network development cost. Furthermore, SOSANETs can be extended and maintained while on-line.

\subsection{The $\pi$-calculus} \label{pi}

The $\pi$-calculus is a simple but powerful model of computation for concurrent systems. It was first published 1992 in "A calculus of mobile processes" \cite{milner1992} by  Robin Milner, Joachim Parrow and David Walker. The word mobility stands here for the movement of links in the virtual space of linked processes \cite{pi}. \\
The $\pi$-calculus has two basic entities: names and processes. Processes are an abstraction of an independent thread of control. Names are the channels (alternative: communication links) processes use to communicate by sending and receiving messages \cite{pi}. The ability to pass channels as data along other channels which allows to model process mobility distinguishes the $\pi$-calculus from preceding process algebras.\\
The used syntax for processes, actions and names is presented in Table \ref{tab:pi}. \\ 

\begin{table}[!ht]
	\centering
	\begin{tabular}{r l l l l l}

  $\pi$ :: &  = $\bar{x}$y &| x(y) &| $\tau$ &&	\\

     P ::&=  \textbf{0} 	&| $\pi$.P &| $P_1$|$P_2$ &| $P_1$ + $P_2$ &| !P | ($\nu$z)P\\  

	\end{tabular}
	\caption{$\pi$-calculus syntax}
	\label{tab:pi}
\end{table}

There are three prefixes expressing the capabilities of actions. The first two represent the sending or receiving of a name. The third stands for a silent transition. The term $\pi$.P denotes that P cannot proceed until action $\pi$ occurred. Respectively, \textbf{$\bar{x}$y.P} is an \textit{output prefix} saying that P can send the name y via channel x and than proceed as P. The \textit{input prefix} \textbf{x(y).P} states that P can receive any name via channel x before it behaves like P. The \textit{unobservable prefix} \textbf{$\tau$}\textbf{.P} can evolve invisible to P \cite{pi}. \textbf{$P_1$|$P_2$} is a \textit{parallel composition} of two processes, while \textbf{$P_1$ + $P_2$} models the \textit{choice} between two variants of behavior. \textbf{!P} is called \textit{replication} denoting an infinite number of instances of P running in parallel and \textbf{0} is an \textit{inert process} that can do nothing. The \textit{restriction} \textbf{($\nu$z)P} ensures  that z is a new channel restricted to P allowing a private interaction between components of P.\\

This short list of constructs is enough to represent all imaginable variations of concurrent behavior.


\subsection{KLAIM}\label{KLAIM}

KLAIM (kernel-language for agent interaction and mobile computing) is an experimental language designed to model Wide Area Networks (WANs) composed out of diverse, mobile components. It was first published 1998 by de Nicola, Ferrari and Pugliese \cite{deNicola1998_klaim}. The language is based on the $\pi$-calculus and the coordination language \textit{Linda} \cite{Gelernter89_Linda}. It extends the expressive $\pi$-calculus with predefined actions for data manipulation as found in Linda. KLAIM allows dynamic channel creation. The communication is asynchronous. Although links can be built on the fly communication partners have to be known in advance.\\
\begin{table}[!ht]
	\centering
	\begin{tabular}{|r l r c r l r|}
\hline
   P : : &  =     		 &  \textsc{ Processes} 	&	&N::&=&\textsc{Nets}		\\

     &  \textbf{nil} 	 & null process 			& & & \textbf{0}      &    empty net			\\
     & | act.P 				 & action prefixing 	&&& $s::^S_{\rho}$P &    single node 	 \\
     
		 & | $P_1$ | $P_2$ & parallel composition &&&$N_1$ || $N_2$  &    net composition 	 \\
		 
		 & | $P_1$ + $P_2$ & choice	 &&&&\\
		 
		 & | X	 						 & process variable		&&&& \\	 
\hline
	\end{tabular}
	\caption[OpenKlaim syntax]{KLAIM syntax for processes and nets}
	\label{tab:process}
\end{table}

In Table \ref{tab:process} the KLAIM syntax for processes is listed. It follows the notations which are commonly used in process algebras and are based on Milner's CCS \cite{milner1989}. \textbf{nil} stands for a process that can not perform any actions. \textit{act.P} states that after the execution of action act the process behaves like P. There are two ways to compose two processes $P_1$ and $P_2$: $P_1$ | $P_2$ stands for a parallel composition, $P_1$ + $P_2$ for a nondeterministic one. Furthermore, a syntax for nets is described. A net can be an empty net \textbf{0}. It can be a single node or the parallel composition of two nets $N_1$ and $N_2$ with a disjoint set of node sites. \\
Following syntactic categories are used from here on:
  \begin{itemize}
	   \item \textit{Sites}: identifier through which a node can be uniquly identified through the network (comparable to IP adresses).
     \item \textit{Localities}: symbolic name for a site allowing to structure programs over distributed environments without knowing their precise allocation. The locality \textbf{self} is used by processes to refer to their execution site. 
		 \item \textit{$\rho$}: is the allocation environment, a finite partial function mapping localities with sites. 
     \item  \textit{Network node}: 4-tuple $s::^S_{\rho}P$, where \textit{s} is the site of the node, \textit{S} the set of sites logged into \textit{s}, $P$ a set of concurrent processes running on \textit{s} and $\rho$ the allocation environment of \textit{s}.
  \end{itemize}

The comunication between processes is supported through multiple distributed \textbf{tuple spaces (TS)}. There is a tuple space on each node. The TS can be seen as a local repository for data and interactions.    \textbf{Tuples (t)} are sequences of information items and are anonymous. They are retrieved from tuple spacing using pattern-matching mechanisms and are manipulated with four predefined actions \textit{act}. These actions are: 
  \begin{itemize}
\item \textbf{out(t)}: produces a tuple by writing it into the TS 
\item \textbf{eval(t)}:  creates new instances of processes for tuple evaluation and writes the results into the TS
\item  \textbf{in(t)}: reads and consumes a tuple from the TS
\item  \textbf{read(t)}: reads tuples non-destructively
\item  \textbf{bind($\ell$,s)}: enhances local allocation environments with new aliases for sites 
\end{itemize}
There is a clear separation between the computational level and the net coordinator/ administrator level in KLAIM. OpenKLAIM \cite{klaim2003}, a dialect of KLAIM, introduces a new category of processes called \textit{NodeCoordinators}. They can be interpretated as kind of superuser managing topology changes within the network. NodeCoordinators can perform privileged actions to create new nodes and establish and/or remove communication channels. These priveledged actions are listed below.
 \begin{itemize}
		\item \textbf{newloc}(s,$\mathbb{C}$)	creates a new node in the network with the site s and installes a NodeCoordinator process $\mathbb{C}$ on the new node. 
	  \item \textbf{login}($\ell$) builts a connection between the performing node and the node with locality $\ell$. 
	  \item \textbf{accept}(s) accept the establishing of a communication channel with the node having the site s. All authorized nodes have to be known in advance. 
	  \item \textbf{logout}($\ell$) removes the link between the action performing node and the node with locality $\ell$.
\end{itemize}

They were introduced to make KLAIM suitable for dealing with open systems when \textit{naming} might not be enough for establishing new connections \cite{klaim2003}. This could be the case when networks routes are affected by restrictions, for example through firewall policies. Still, there is a limitation resulting out of the definitions of these actions. The syntax of the privileged action \textit{accept} shows that all future connection partners of one node have to be known at design level. This limits the capability of KLAIM to handle the addition of completly unexpected nodes restricting the evolvement possibilities of the SOSANET.\\
A further mechanism for access control in KLAIM is a capability-based type system. It provides information about permissions for the download and consumption of tuples, the activation of processes and the creation of new nodes (more detailed information can be found in \cite{nicola2000}).\\
Additionally KLAIM provides a programming language X-KLAIM (eXtended KLAIM)\cite{klaim2003} with a high-level syntax for processes providing variable declarations, enriched operations, assignments, conditionals, sequential and iterative process composition. The implementation is based on KLAVA, a Java package providing the run-time system for X-KLAIM operations and a compiler for the X-KLAIM to Java translation. The hierarchical model of KLAIM is preserved providing all primitives for the handling of dynamic node connectivity.\\

We introduced SOSANETs in this section describing its advantages over classic SANET designs. Furthermore, we presented the $\pi$-calculus and KLAIM pointing out its benefits for our goal and the limitations. In the following sections we will apply the described actions and processes of KLAIM to model service-oriented sensor-actuator networks.

\section{A mobile SOSANET: The Driver Assistant System}\label{model}
\begin{figure}[h!]%
\begin{center}
\includegraphics[width=1\textwidth]{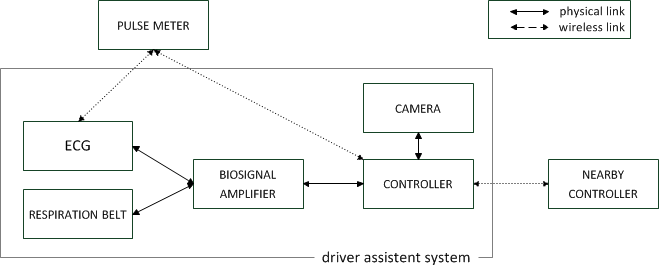}%
\caption{Use case scenario: a driver assistant system.}%
\label{fig:das}%
\end{center}
\end{figure}

We are working on a use case to evaluate the modeling power of our chosen $\pi$-calculus variant building a test surrounding for future work on network behavior analysis and verification. Our case study is a driver assistant system. It is illustrated in Figure \ref{fig:das}.\\

The system measures different vital signs of the driver and is able to react to unexpected events like heart attacks by sending an alarm and performing an emergency brake. The inner network is composed out of five wired components: a controller, and signal amplifier, a camera, a respiration belt and an electrocardiogram (ECG). The camera observes the head and eye movement, ECG and the respiration belt measure the breathing and the heart beat of the driver. The data evaluation and the activation of an alarm are part of the controller functionality. We assume that the driver also has a assistant system at his apartment securing his health status at home. This system requires a constant application of a portable pulse meter. It becomes part of the driver assistant network when the bearer enters the car by building wireless communication channels with the ECG and the controller. The data of the pulse meter becomes redundant when the ECG starts its measurements. In this case, the pulse meter can switch into a battery saving mode. One or more additional communication links can be established on-line in case of an emergency that requires information exchange with surrounding traffic participants. While the addition of the pulse meter node is an expected one, the later ones are not.\\

We focus in our modeling examples for clarity on four nodes and their interaction: the controller unit (cu), the amplifier (ampl) the ECG (ecg) and the pulse meter (pm) as illustrated in Figure \ref{fig:soa}. Hence, we have:\\
\setlength{\abovedisplayskip}{0.1cm}
\setlength{\belowdisplayskip}{0.4cm}
\begin{eqnarray*}
N&=&s_{cu}::^{\{s_{ampl},s_{pm}\}}_{s_{cu}/self,s_{ampl}/\ell_{ampl},s_{pm}/\ell_{pm}}P_{cu}~||~s_{ampl}::^{\{s_{cu},s_{ecg}\}}_{s_{ampl}/self,s_{cu}/\ell_{cu},s_{ecg}/\ell_{ecg}}P_{ampl}~||\\
&& s_{ecg}::^{\{s_{ampl},s_{pm}\}}_{s_{ecg}/self,s_{ampl}/\ell_{ampl},s_{pm}/\ell_{pm}}P_{ecg}~||~s_{pm}::^{\{s_{cu},s_{ecg}\}}_{s_{pm}/self,s_{cu}/\ell_{cu},s_{ecg}/\ell_{ecg}}P_{pm}
\end{eqnarray*}

This network description states that the network under observation consists out of four nodes which have the unique network identifiers (sites) $s_{cu}$, $s_{ecg}$, $s_{ampl}$ and $s_{pm}$. The controller node is connected with the amplifier node and the pulse meter node mapping the respective sites with the associated localities. The set $P_{cu}$ contains all processes which run concurrently on the node. The formula modeling the other nodes can be interpreted equivalently.  

\subsection*{Modeling Topology Changes}

We have to differentiate between four kinds of possible topology changes that can occur in a SOSANET.\\

The first one is the creation of a \textit{new link between two familiar network nodes}. That could be the establishment of a new communication channel between already existing members of the network. Or it is the connection between a node which enters the network and a member of the network that is expecting the new node. This behavior can be modeled using the privileged NodeCoordinator actions login($\ell$) and accept(s) in combination with newloc(s,$\mathbb{C}$).\\
\setlength{\abovedisplayskip}{0.1cm}
\setlength{\belowdisplayskip}{0.4cm}
\begin{eqnarray*}
pm_{new} &=&newloc(s_pm,\mathbb{C}_{pm}).\mathbb{C}'_{pm} ~| ~pm_{dorm}\\
\mathbb{C}'_{pm}&=&login(\ell_{cu}).login(\ell_{ecg}).C''_{pm}\\
\mathbb{C}_{cont} &=&\sum_{s_i} accept(s_i).\mathbb{C}_{cont} + ...\\
\mathbb{C}_{ecg} &=&\sum_{s_i} accept(s_i).\mathbb{C}_{ecg} + ...\\
...
\end{eqnarray*}

In the example above the entering of the pulse meter into the driver assistant network is described. First a new node with the network identifier $s_{pm}$ is created and a NodeCoordinator instance is installed. The behavior of the node splits into two parallel components. One is the pulse meter in energy saving mode and the other a NodeCoordinator process. The NodeCoordinator tries to built new communication channels between the pulse meter and the ECG and  between the pulse meter and the controller node. The corresponding NodeCoordinator processes have to synchronize to create the desired link. Furthermore, the site of pm has to be known by the accepting partner to complete the task.\\

Another possibility is that a \textit{node with an unfamiliar site tries to enter the network}. This would be for example a controller node which belongs to a network installed in another car. The node wants to inform the first system that an emergency occurred in its driver cabin and that it is going to execute an emergency brake within the next moments. There is no possibility in KLAIM to model this scenario. We are only able to describe the creation of the new node and its attempts to open communication channels. An \textit{accept} can not be processed due to the unknown site of the connecting node.\\

There is also the possibility to model the \textit{controlled removal of a communication link} between two nodes.\\

\setlength{\abovedisplayskip}{0.1cm}
\setlength{\belowdisplayskip}{0.4cm}
\begin{eqnarray*}
\mathbb{C}_{pm}&=& logout(\ell_{cu}).logout(\ell_{ecg}).\mathbb{C}_{pm}~ | ... \\
...
\end{eqnarray*}

The pulse meter leaves the network by removing the previously established links. This happens when the driver is leaving the car after he reached his destination.\\

The last kind of topology changes we like to consider is the \textit{spontaneous linkage failure}. These failures are only discoverable using special algorithms or test signals which check the functionality of the observed channel. The actions we used above are not appliable to model this behavior. 

\subsection*{Modeling Service Functionalities}\label{subSOA}

\begin{figure}[h!]%
\begin{center}
\includegraphics[width=1\textwidth]{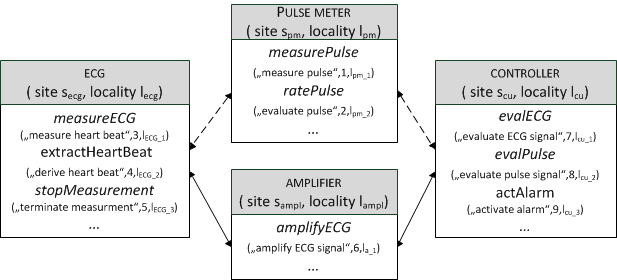}%
\caption{The reduced use case with a selection of offered services displayed at each node.}%
\label{fig:soa}%
\end{center}
\end{figure}

There are three basic operations in service oriented architectures needed for the interaction between service providers and service users: \textit{service publishing, service discovery} and \textit{service request}. In general, the \textit{service publishing} function publishes the description of a service, its location and the associated service identifier into a central service registry. There is no such central registry in service-oriented sensor-actuator networks due to limited resources on each node. Furthermore, the ability of the network to change its topology ad hoc could lead to the disconnection of the central "`registry node"' immobilizing the whole network.\\
Here, each node has a local registry storing informations about its own services and routes to other network services. This is implemented by writing 3-tuples of the form ("description", id, locality) into the local tuple space. If the third tuple entry is \textbf{self}, it states, the service is installed on the node itself. Every other entry refers to a connected, neighboring node which has further information on the service locality in its tuple space. Figure \ref{fig:soa} illustrates a service view on the network. Only a selection of services is published to keep the figure clear.\\

A formal notation of this function, here we observe the publishing of the \textit{measure pulse} service, can be seen below. We focus on the exact execution order of each involved action and process omitting the description of possible process behavior alternatives for clarity reasons.\\
\setlength{\abovedisplayskip}{0.1cm}
\setlength{\belowdisplayskip}{0.4cm}
\begin{eqnarray*}
pm & = & out("measure~pulse",1,self)@self.out("measure~pulse",1,\ell_{pm})@\ell_{ECG}.\\
&&out("measure~pulse",1,\ell_{pm})@\ell_{cu}.pm\\
ECG & = &  out(in(!description,!service_{id},!loc))@self.\\
&&out(description,service_{id},loc)@\ell_{ampl}.ECG\\
ampl & = & ...\\
...
\end{eqnarray*}

First, the pulse meter publishes the existence of the service by writing the new tuple ("measure pulse",1,self) into its tuple space. Afterwards it forwards the information to all directly connected nodes. Then it behaves itself again like before. \\ The ECG node waits for invocation, receives the information and stores it in its own local tuple space before sending the received data to all neighboring nodes. This behavior is executed at each node in the network. The publishing process ends when a node receives a tuple which matches an existing one (not described above).\\

The second SOA operation, the \textit{service discovery}, searches for an available service within the network. The search key is either the service id or a service description. The first one is used, if we are looking for a specific service. If we are satisfied with any service providing the needed functionality, we use a the second search key option. The search is done using the mobile process \textit{discover}.\\First, the controller node asks for the execution of the process \textit{discover}. The process travels after invocation dynamically between nodes searching through the local tuple spaces for the locality of a service, which can measure the pulse (search key: "measure pulse"). Here the key and the locally saved tuples are compared with pattern matching mechanisms. The controller node splits its behavior into two components, $cu_{1}$ waiting for a response and $cu_{3}$ proceeding. This decouples other functionalities of the node from the search activities avoiding a blockage of all available processes.\\  
\setlength{\abovedisplayskip}{0.1cm}
\setlength{\belowdisplayskip}{0.4cm}
\begin{eqnarray*}
cu & = & newloc(\ell_1).eval(discover("measure~pulse",id,\ell_1))@\ell_{cu}.\\
&&((in(!loc)@\ell_{1}.cu_{1})~ |~ cu_{3})\\
...
\end{eqnarray*}
\textit{Discover } can match two tuples. The first one captures the locality information of the inquired service, back-propagates it through the network using the locality $\ell_1$ and terminates the process \textit{discover}. The other tuple captures the locality where the search has to be repeated.  \\
\setlength{\abovedisplayskip}{0.1cm}
\setlength{\belowdisplayskip}{0.4cm}
\begin{eqnarray*}
discover(description,id,loc)&=& read("measure~pulse",id,!loc)@self.out(loc)@\ell_1.nil\\
&+& read("measure~pulse",id,\ell_1')@self.\\
&&eval(discover("measure~pulse",id,\ell_1))@\ell_1'.nil\\
...
\end{eqnarray*}

The execution of a services can be requested after discovering the service location. This in done using the \textit{service request} operation. In our example the controller node wants to invoke a measurement of the pulse of the driver to check his health status. This means it requests an execution of the \textit{measurePulse} service on the pulse meter node. \\
The instance of the controller that waited for the location of the queried service, sends a service request using the inquired locality information. Then it waits for the a transmission of the requested data. It behaves like $cu_2$ after receiving the data. The pulse meter node waits simultaneously for invocation. After receiving the request it splits its behavior into two component. One instance executes the requested service and terminates after sending the inquired data, the other behaves as before.\\
\setlength{\abovedisplayskip}{0.1cm}
\setlength{\belowdisplayskip}{0.4cm}
\begin{eqnarray*}
cu_1 &=& newloc(\ell_2).out("measurePulse",\ell_2)@loc.in(!pulseData)@\ell_2.cu_2\\
pm &= &in(!reqService,!loc).\\
&&(<reqService>.out(pulseData)@loc.nil | pm)\\
cu_2 &=& <evalPulse>.out(result)@self.cu_3\\
cu_3 &=& ...\\
...
\end{eqnarray*}
The process $cu_2$ uses its own functionality \textit{evalPulse} to evaluate the received data and to forward the result to process $cu_3$ which reacts appropriately (not noted).\\

As shown in this section, the given primitives and actions of KLAIM are expressive enough to formally describe features and behavior of SOSANETs. It is possible to formalize single node and complete network structures including available processes, communication channels and knowledge about dependencies between sites and localities. We are able to model process behavior, topology changes (the addition of expected nodes or the establishment and removal of connections) and the three central operations of service oriented architectures.

\section{Conclusion and Future Work}\label{conclusion}

We have illustrated the need for the analysis and verification of the behavior of service-oriented sensor-actuator networks applied in the medical domain. We could show that our chosen $\pi$-calculus variant KLAIM is suitable for modeling SOSA-NETs under the restriction that all topology changes are known in advance at design time. Furthermore, we could see that the provided actions for tuple manipulation are beneficial for the description of service-oriented routines. We have evaluated the applicability of KLAIM as formal modeling language for SOSANETs by modeling a driver assistant system. Furthermore, we identified the need to extend the language. Actions are needed, which are able to represent the addition of unexpected nodes. Without that, the interaction of unrelated networks cannot be analyzed limiting the constructional power of SOSANETs.\\
Our future work will focus on the described extension and on methods for the analysis and verification of network behavior. After finishing these goals we will be able to present a formal design framework for service-oriented sensor-actuator networks.\\

\textbf{Acknowledgment.} We thank the Geriatric Center of the Charit\'{e} Berlin for familiarizing us with a closed driver assistant system which builds the basis of our case study.

\nocite{*}
\bibliographystyle{eptcs}
\bibliography{literature}

\begin{thebibliography}{10}
\providecommand{\bibitemdeclare}[2]{}
\providecommand{\surnamestart}{}
\providecommand{\surnameend}{}
\providecommand{\urlprefix}{Available at }
\providecommand{\url}[1]{\texttt{#1}}
\providecommand{\href}[2]{\texttt{#2}}
\providecommand{\urlalt}[2]{\href{#1}{#2}}
\providecommand{\doi}[1]{doi:\urlalt{http://dx.doi.org/#1}{#1}}
\providecommand{\bibinfo}[2]{#2}

\bibitemdeclare{article}{Akyildiz2004}
\bibitem{Akyildiz2004}
\bibinfo{author}{Ian~F. \surnamestart Akyildiz\surnameend} \&
  \bibinfo{author}{Ismail~H. \surnamestart Kasimoglu\surnameend}
  (\bibinfo{year}{2004}): \emph{\bibinfo{title}{Wireless sensor and actor
  networks: research challenges}}.
\newblock {\sl \bibinfo{journal}{Ad Hoc Networks}}
  \bibinfo{volume}{2}(\bibinfo{number}{4}), pp. \bibinfo{pages}{351 -- 367},
  \doi{10.1016/j.adhoc.2004.04.003}.
\newblock
  \urlprefix\url{http://www.sciencedirect.com/science/article/pii/S1570870504000319}.

\bibitemdeclare{inproceedings}{balasubramaniam2008}
\bibitem{balasubramaniam2008}
\bibinfo{author}{D.~\surnamestart Balasubramaniam\surnameend},
  \bibinfo{author}{A.~\surnamestart Dearle\surnameend} \&
  \bibinfo{author}{R.~\surnamestart Morrison\surnameend}
  (\bibinfo{year}{2008}): \emph{\bibinfo{title}{A composition-based approach to
  the construction and dynamic reconfiguration of wireless sensor network
  applications}}.
\newblock In: {\sl \bibinfo{booktitle}{Software Composition}},
  \bibinfo{organization}{Springer}, pp. \bibinfo{pages}{206--214},
  \doi{10.1007/978-3-540-78789-1\_16}.

\bibitemdeclare{article}{klaim2003}
\bibitem{klaim2003}
\bibinfo{author}{Lorenzo \surnamestart Bettini\surnameend},
  \bibinfo{author}{Viviana \surnamestart Bono\surnameend},
  \bibinfo{author}{Rocco~De \surnamestart Nicola\surnameend},
  \bibinfo{author}{Gianluigi \surnamestart Ferrari\surnameend},
  \bibinfo{author}{Daniele \surnamestart Gorla\surnameend},
  \bibinfo{author}{Michele \surnamestart Loreti\surnameend},
  \bibinfo{author}{Eugenio \surnamestart Moggi\surnameend},
  \bibinfo{author}{Rosario \surnamestart Pugliese\surnameend},
  \bibinfo{author}{Emilio \surnamestart Tuosto\surnameend} \&
  \bibinfo{author}{Betti \surnamestart Venneri\surnameend}
  (\bibinfo{year}{2003}): \emph{\bibinfo{title}{The Klaim Project: Theory and
  Practice}}.
\newblock {\sl \bibinfo{journal}{Global Computing: Programming Environments,
  Languages, Security and Analysis of Systems}} \bibinfo{volume}{2874}, pp.
  \bibinfo{pages}{88--150}, \doi{10.1007/978-3-540-40042-4\_4}.

\bibitemdeclare{article}{Bucur2011}
\bibitem{Bucur2011}
\bibinfo{author}{Doina \surnamestart Bucur\surnameend} \&
  \bibinfo{author}{Marta \surnamestart Kwiatkowska\surnameend}
  (\bibinfo{year}{2011}): \emph{\bibinfo{title}{On software verification for
  sensor nodes}}.
\newblock {\sl \bibinfo{journal}{Journal of Systems and Software}}
  \bibinfo{volume}{84}(\bibinfo{number}{10}), pp. \bibinfo{pages}{1693 --
  1707}, \doi{10.1016/j.jss.2011.04.054}.
\newblock
  \urlprefix\url{http://www.sciencedirect.com/science/article/pii/S0164121211001051}.

\bibitemdeclare{article}{Gelernter89_Linda}
\bibitem{Gelernter89_Linda}
\bibinfo{author}{Nicholas \surnamestart Carriero\surnameend} \&
  \bibinfo{author}{David \surnamestart Gelernter\surnameend}
  (\bibinfo{year}{1989}): \emph{\bibinfo{title}{Linda in context}}.
\newblock {\sl \bibinfo{journal}{Commun. ACM}}
  \bibinfo{volume}{32}(\bibinfo{number}{4}), pp. \bibinfo{pages}{444--458},
  \doi{10.1145/63334.63337}.

\bibitemdeclare{article}{deNicola1998_klaim}
\bibitem{deNicola1998_klaim}
\bibinfo{author}{R.~\surnamestart De~Nicola\surnameend}, \bibinfo{author}{G.L.
  \surnamestart Ferrari\surnameend} \& \bibinfo{author}{R.~\surnamestart
  Pugliese\surnameend} (\bibinfo{year}{1998}): \emph{\bibinfo{title}{KLAIM: a
  kernel language for agents interaction and mobility}}.
\newblock {\sl \bibinfo{journal}{Software Engineering, IEEE Transactions on}}
  \bibinfo{volume}{24}(\bibinfo{number}{5}), pp. \bibinfo{pages}{315 --330},
  \doi{10.1109/32.685256}.

\bibitemdeclare{article}{nicola2000}
\bibitem{nicola2000}
\bibinfo{author}{R.~\surnamestart De~Nicola\surnameend}, \bibinfo{author}{G.L.
  \surnamestart Ferrari\surnameend} \& \bibinfo{author}{R.~\surnamestart
  Pugliese\surnameend} (\bibinfo{year}{2000}):
  \emph{\bibinfo{title}{Programming Access Control: The Klaim Experience}}.
\newblock {\sl \bibinfo{journal}{CONCUR 2000—Concurrency Theory}}, pp.
  \bibinfo{pages}{48--65}.

\bibitemdeclare{article}{KLAIM2006}
\bibitem{KLAIM2006}
\bibinfo{author}{R.~\surnamestart De~Nicola\surnameend},
  \bibinfo{author}{D.~\surnamestart Gorla\surnameend} \&
  \bibinfo{author}{R.~\surnamestart Pugliese\surnameend}
  (\bibinfo{year}{2006}): \emph{\bibinfo{title}{On the expressive power of
  KLAIM-based calculi}}.
\newblock {\sl \bibinfo{journal}{Theoretical Computer Science}}
  \bibinfo{volume}{356}(\bibinfo{number}{3}), pp. \bibinfo{pages}{387--421},
  \doi{10.1016/j.tcs.2006.02.007}.

\bibitemdeclare{inproceedings}{dearle2008}
\bibitem{dearle2008}
\bibinfo{author}{A.~\surnamestart Dearle\surnameend},
  \bibinfo{author}{D.~\surnamestart Balasubramaniam\surnameend},
  \bibinfo{author}{J.~\surnamestart Lewis\surnameend} \&
  \bibinfo{author}{R.~\surnamestart Morrison\surnameend}
  (\bibinfo{year}{2008}): \emph{\bibinfo{title}{A component-based model and
  language for wireless sensor network applications}}.
\newblock In: {\sl \bibinfo{booktitle}{Computer Software and Applications,
  2008. COMPSAC'08. 32nd Annual IEEE International}},
  \bibinfo{organization}{IEEE}, pp. \bibinfo{pages}{1303--1308},
  \doi{10.1109/COMPSAC.2008.151}.

\bibitemdeclare{inproceedings}{Diaz2006}
\bibitem{Diaz2006}
\bibinfo{author}{M.~\surnamestart Diaz\surnameend},
  \bibinfo{author}{D.~\surnamestart Garrido\surnameend},
  \bibinfo{author}{L.~\surnamestart Llopis\surnameend},
  \bibinfo{author}{B.~\surnamestart Rubio\surnameend} \& \bibinfo{author}{J.M.
  \surnamestart Troya\surnameend} (\bibinfo{year}{2006}):
  \emph{\bibinfo{title}{A Component Framework for Wireless Sensor and Actor
  Networks}}.
\newblock In: {\sl \bibinfo{booktitle}{Emerging Technologies and Factory
  Automation, 2006. ETFA '06. IEEE Conference on}}, pp. \bibinfo{pages}{300
  --307}, \doi{10.1109/ETFA.2006.355382}.

\bibitemdeclare{inproceedings}{dunkels2004contiki}
\bibitem{dunkels2004contiki}
\bibinfo{author}{A.~\surnamestart Dunkels\surnameend},
  \bibinfo{author}{B.~\surnamestart Gronvall\surnameend} \&
  \bibinfo{author}{T.~\surnamestart Voigt\surnameend} (\bibinfo{year}{2004}):
  \emph{\bibinfo{title}{Contiki-a lightweight and flexible operating system for
  tiny networked sensors}}.
\newblock In: {\sl \bibinfo{booktitle}{Local Computer Networks, 2004. 29th
  Annual IEEE International Conference on}}, \bibinfo{organization}{IEEE}, pp.
  \bibinfo{pages}{455--462}, \doi{10.1109/LCN.2004.38}.

\bibitemdeclare{article}{francalanza2008}
\bibitem{francalanza2008}
\bibinfo{author}{Adrian \surnamestart Francalanza\surnameend} \&
  \bibinfo{author}{Matthew \surnamestart Hennessy\surnameend}
  (\bibinfo{year}{2008}): \emph{\bibinfo{title}{A theory of system behaviour in
  the presence of node and link failure}}.
\newblock {\sl \bibinfo{journal}{Information and Computation}}
  \bibinfo{volume}{206}(\bibinfo{number}{6}), pp. \bibinfo{pages}{711 -- 759},
  \doi{10.1016/j.ic.2007.12.002}.
\newblock
  \urlprefix\url{http://www.sciencedirect.com/science/article/pii/S0890540108000023}.

\bibitemdeclare{inproceedings}{conf/sensornets/HarveyDLS12}
\bibitem{conf/sensornets/HarveyDLS12}
\bibinfo{author}{Paul \surnamestart Harvey\surnameend}, \bibinfo{author}{Alan
  \surnamestart Dearle\surnameend}, \bibinfo{author}{Jonathan \surnamestart
  Lewis\surnameend} \& \bibinfo{author}{Joseph~S. \surnamestart
  Sventek\surnameend} (\bibinfo{year}{2012}): \emph{\bibinfo{title}{Channel and
  Active Component Abstractions for WSN Programming - A Language Model with
  Operating System Support.}}
\newblock In \bibinfo{editor}{Marten \surnamestart van Sinderen\surnameend},
  \bibinfo{editor}{Octavian \surnamestart Postolache\surnameend} \&
  \bibinfo{editor}{César \surnamestart Benavente-Peces\surnameend}, editors:
  {\sl \bibinfo{booktitle}{SENSORNETS}}, \bibinfo{publisher}{SciTePress}, pp.
  \bibinfo{pages}{35--44}.
\newblock
  \urlprefix\url{http://dblp.uni-trier.de/db/conf/sensornets/sensornets2012.html#HarveyDLS12}.

\bibitemdeclare{inproceedings}{Jacoub2011}
\bibitem{Jacoub2011}
\bibinfo{author}{J.~\surnamestart Khalil~Jacoub\surnameend},
  \bibinfo{author}{R.~\surnamestart Liscano\surnameend} \&
  \bibinfo{author}{J.~\surnamestart Bradbury\surnameend}
  (\bibinfo{year}{2011}): \emph{\bibinfo{title}{A Survey of Modeling Techniques
  for Wireless Sensor Networks}}.
\newblock In: {\sl \bibinfo{booktitle}{SENSORCOMM 2011, The Fifth International
  Conference on Sensor Technologies and Applications}}, pp.
  \bibinfo{pages}{103--109}.

\bibitemdeclare{article}{Jacoub2012}
\bibitem{Jacoub2012}
\bibinfo{author}{J.~\surnamestart Khalil~Jacoub\surnameend},
  \bibinfo{author}{R.~\surnamestart Liscano\surnameend} \&
  \bibinfo{author}{J.~\surnamestart Bradbury\surnameend}
  (\bibinfo{year}{2012}): \emph{\bibinfo{title}{Assessment of Software Modeling
  Techniques for Wireless Sensor Networks: A Survey}}.
\newblock {\sl \bibinfo{journal}{Sensors and Transducers}}
  \bibinfo{volume}{14-2}, pp. \bibinfo{pages}{18--46}.

\bibitemdeclare{inproceedings}{greatduckisland}
\bibitem{greatduckisland}
\bibinfo{author}{Alan~M. \surnamestart Mainwaring\surnameend},
  \bibinfo{author}{David~E. \surnamestart Culler\surnameend},
  \bibinfo{author}{Joseph \surnamestart Polastre\surnameend},
  \bibinfo{author}{Robert \surnamestart Szewczyk\surnameend} \&
  \bibinfo{author}{John \surnamestart Anderson\surnameend}
  (\bibinfo{year}{2002}): \emph{\bibinfo{title}{Wireless sensor networks for
  habitat monitoring}}.
\newblock In: {\sl \bibinfo{booktitle}{WSNA}}, pp. \bibinfo{pages}{88--97}.
\newblock \urlprefix\url{http://doi.acm.org/10.1145/570738.570751}.

\bibitemdeclare{article}{martinez2011}
\bibitem{martinez2011}
\bibinfo{author}{Diego \surnamestart Martinez\surnameend},
  \bibinfo{author}{Apolinar \surnamestart Gonzalez\surnameend},
  \bibinfo{author}{Francisco \surnamestart Blanes\surnameend},
  \bibinfo{author}{Raul \surnamestart Aquino\surnameend}, \bibinfo{author}{Jose
  \surnamestart Simo\surnameend} \& \bibinfo{author}{Alfons \surnamestart
  Crespo\surnameend} (\bibinfo{year}{2011}): \emph{\bibinfo{title}{Formal
  Specification and Design Techniques for Wireless Sensor and Actuator
  Networks}}.
\newblock {\sl \bibinfo{journal}{Sensors}}
  \bibinfo{volume}{11}(\bibinfo{number}{1}), pp. \bibinfo{pages}{1059--1077},
  \doi{10.3390/s110101059}.
\newblock \urlprefix\url{http://www.mdpi.com/1424-8220/11/1/1059}.

\bibitemdeclare{inproceedings}{DBLP:conf/sutc/MeshkovaROJM08}
\bibitem{DBLP:conf/sutc/MeshkovaROJM08}
\bibinfo{author}{Elena \surnamestart Meshkova\surnameend},
  \bibinfo{author}{Janne \surnamestart Riihij{\"a}rvi\surnameend},
  \bibinfo{author}{Frank \surnamestart Oldewurtel\surnameend},
  \bibinfo{author}{Christine \surnamestart Jardak\surnameend} \&
  \bibinfo{author}{Petri \surnamestart M{\"a}h{\"o}nen\surnameend}
  (\bibinfo{year}{2008}): \emph{\bibinfo{title}{Service-Oriented Design
  Methodology for Wireless Sensor Networks: A View through Case Studies}}.
\newblock In: {\sl \bibinfo{booktitle}{SUTC}}, pp. \bibinfo{pages}{146--153}.
\newblock
  \urlprefix\url{http://doi.ieeecomputersociety.org/10.1109/SUTC.2008.43}.

\bibitemdeclare{book}{milner1989}
\bibitem{milner1989}
\bibinfo{author}{R.~\surnamestart Milner\surnameend} (\bibinfo{year}{1989}):
  \emph{\bibinfo{title}{Communication and concurrency}}.
\newblock \bibinfo{publisher}{Prentice-Hall, Inc.}

\bibitemdeclare{book}{pi}
\bibitem{pi}
\bibinfo{author}{Robin \surnamestart Milner\surnameend} (\bibinfo{year}{1999}):
  \emph{\bibinfo{title}{Communicating and Mobile Systems: The pi-Calculus}}.
\newblock \bibinfo{publisher}{Cambridge University Press}.

\bibitemdeclare{article}{milner1992}
\bibitem{milner1992}
\bibinfo{author}{Robin \surnamestart Milner\surnameend},
  \bibinfo{author}{Joachim \surnamestart Parrow\surnameend} \&
  \bibinfo{author}{David \surnamestart Walker\surnameend}
  (\bibinfo{year}{1992}): \emph{\bibinfo{title}{A calculus of mobile processes,
  I}}.
\newblock {\sl \bibinfo{journal}{Inf. Comput.}}
  \bibinfo{volume}{100}(\bibinfo{number}{1}), pp. \bibinfo{pages}{1--40},
  \doi{10.1016/0890-5401(92)90008-4}.

\bibitemdeclare{inproceedings}{oleshchuk2003}
\bibitem{oleshchuk2003}
\bibinfo{author}{V.A. \surnamestart Oleshchuk\surnameend}
  (\bibinfo{year}{2003}): \emph{\bibinfo{title}{Ad-hoc sensor networks:
  modeling, specification and verification}}.
\newblock In: {\sl \bibinfo{booktitle}{Intelligent Data Acquisition and
  Advanced Computing Systems: Technology and Applications, 2003. Proceedings of
  the Second IEEE International Workshop on}}, \bibinfo{organization}{IEEE},
  pp. \bibinfo{pages}{76--79}, \doi{10.1109/IDAACS.2003.1249521}.

\bibitemdeclare{inproceedings}{Papazoglou2003}
\bibitem{Papazoglou2003}
\bibinfo{author}{M.P. \surnamestart Papazoglou\surnameend}
  (\bibinfo{year}{2003}): \emph{\bibinfo{title}{Service-oriented computing:
  concepts, characteristics and directions}}.
\newblock In: {\sl \bibinfo{booktitle}{Web Information Systems Engineering,
  2003. WISE 2003. Proceedings of the Fourth International Conference on}}, pp.
  \bibinfo{pages}{3 -- 12}, \doi{10.1109/WISE.2003.1254461}.

\bibitemdeclare{inproceedings}{Patrignani2012}
\bibitem{Patrignani2012}
\bibinfo{author}{M.~\surnamestart Patrignani\surnameend},
  \bibinfo{author}{N.~\surnamestart Matthys\surnameend},
  \bibinfo{author}{J.~\surnamestart Proenca\surnameend},
  \bibinfo{author}{D.~\surnamestart Hughes\surnameend} \&
  \bibinfo{author}{D.~\surnamestart Clarke\surnameend} (\bibinfo{year}{2012}):
  \emph{\bibinfo{title}{Formal analysis of policies in wireless sensor network
  applications}}.
\newblock In: {\sl \bibinfo{booktitle}{Software Engineering for Sensor Network
  Applications (SESENA), 2012 Third International Workshop on}}, pp.
  \bibinfo{pages}{15 --21}, \doi{10.1109/SESENA.2012.6225728}.

\bibitemdeclare{article}{Rezgui2007SOSANET}
\bibitem{Rezgui2007SOSANET}
\bibinfo{author}{Abdelmounaam \surnamestart Rezgui\surnameend} \&
  \bibinfo{author}{Mohamed \surnamestart Eltoweissy\surnameend}
  (\bibinfo{year}{2007}): \emph{\bibinfo{title}{Service-oriented
  sensor-actuator networks: Promises, challenges, and the road ahead}}.
\newblock {\sl \bibinfo{journal}{Computer Communications}}
  \bibinfo{volume}{30}(\bibinfo{number}{13}), pp. \bibinfo{pages}{2627--2648}.
\newblock \urlprefix\url{http://dx.doi.org/10.1016/j.comcom.2007.05.036}.

\bibitemdeclare{article}{sharma2009}
\bibitem{sharma2009}
\bibinfo{author}{O.~\surnamestart Sharma\surnameend},
  \bibinfo{author}{J.~\surnamestart Lewis\surnameend},
  \bibinfo{author}{A.~\surnamestart Miller\surnameend},
  \bibinfo{author}{A.~\surnamestart Dearle\surnameend},
  \bibinfo{author}{D.~\surnamestart Balasubramaniam\surnameend},
  \bibinfo{author}{R.~\surnamestart Morrison\surnameend} \&
  \bibinfo{author}{J.~\surnamestart Sventek\surnameend} (\bibinfo{year}{2009}):
  \emph{\bibinfo{title}{Towards verifying correctness of wireless sensor
  network applications using Insense and SPIN}}.
\newblock {\sl \bibinfo{journal}{Model Checking Software}}
  \bibinfo{volume}{5578}, pp. \bibinfo{pages}{223--240},
  \doi{10.1007/978-3-642-02652-2\_19}.

\end{thebibliography}
\end{document}